\documentclass[twocolumn,prd,floatfix,preprintnumbers,nofootinbib,superscriptaddress]{revtex4-2}

\usepackage{bm}
\usepackage{times}
\usepackage{amssymb,amsbsy,amsmath,amsfonts}
\usepackage{graphicx}
\usepackage{float}
\usepackage{color}
\usepackage{morefloats}
\usepackage{rotating}
\usepackage{srcltx}
\usepackage{slashed}
\usepackage{subfigure}
\usepackage{multirow,diagbox}
\usepackage{verbatim}
\usepackage{tabularx}
\usepackage{tabularray}
\usepackage[outdir=./]{epstopdf}
\usepackage{epsfig}
\usepackage[normalem]{ulem}
\usepackage{hyperref}
\usepackage{tikz}
\usepackage{braket}
\hypersetup{colorlinks=true,linkcolor=blue,citecolor=blue,menucolor=black,urlcolor=black,filecolor=blue}
\usepackage{balance}

\newcommand\bkthree[3]{\langle \vec{#1} \left| #2 \right| \vec{#3}\rangle}

\newcommand\cuti[2]{\theta(\Lambda_{#1}-#2)}

\newcommand\pli[2]{#2^{l_{#1}}}

\newcommand\rmd[0]{\mathrm{d}}
\newcommand\rmi[0]{\mathrm{i}}
\newcommand\rme[0]{\mathrm{e}}

\begin{document}

\title{Femtoscopic correlation functions for general partial waves: Application to the $\Lambda(1520)$ resonance}

\author{Si-Wei Liu}~\email{liusiwei@impcas.ac.cn}
 \affiliation{State Key Laboratory of Heavy Ion Science and Technology, Institute of Modern Physics, Chinese Academy of Sciences, Lanzhou 730000, China} 
\affiliation{School of Nuclear Sciences and Technology, University of Chinese Academy of Sciences, Beijing 101408, China}
\author{Ju-Jun Xie}~\email{xiejujun@impcas.ac.cn}
\affiliation{State Key Laboratory of Heavy Ion Science and Technology, Institute of Modern Physics, Chinese Academy of Sciences, Lanzhou 730000, China} 
\affiliation{School of Nuclear Sciences and Technology, University of Chinese Academy of Sciences, Beijing 101408, China}
\affiliation{Southern Center for Nuclear-Science Theory (SCNT), Institute of Modern Physics, Chinese Academy of Sciences, Huizhou 516000, China}
\date{\today}

\begin{abstract}

The femtoscopic correlation function has been established in recent years as a high-precision tool for investigating hadron-hadron interactions and exotic states, providing stringent constraints on the dynamics of low-energy strong interactions. However, current research has been predominantly focused on the $s$-wave interaction between hadrons, while studies of higher partial waves remain scarce. We present a general analytical expression for the femtoscopic correlation function in an arbitrary partial wave using the Lippmann-Schwinger equation. This formalism is applied to constrain the $d$-wave $K^-p$ scattering through a combined study of the $K^-p$ correlation function and the $D_{03}$ scattering amplitude of $\bar K N \to \bar K N$ and $\bar K N \to \pi\Sigma$ processes, from which the properties of $\Lambda(1520)$ are extracted and found to be in good agreement with the experimental results. These findings demonstrate the feasibility of determining dynamics between hadrons through femtoscopic correlation functions and scattering amplitudes with higher partial waves.

\end{abstract}

\maketitle

\section{Introduction} \label{sec:Introduction}

The non-perturbative nature of quantum chromodynamics (QCD) at low-energy presents significant challenges for analytically studying the strong interaction. Therefore, the study of hadron-hadron interactions is regarded as a key approach for the phenomenological understanding of QCD in low energy region. Traditionally, the investigations of hadron-hadron interactions have been based on the measurement of invariant mass spectra and scattering experiments~\cite{Wang:2024jyk,Xie:2024wbd,Xie:2005sb,Jia:2023upb,BESIII:2020nme,Liu:2023jwo,Dai:2025hvo}. Recently, a new complementary approach has been developed through the femtoscopic correlation function in heavy-ion collisions and $pp$ collision experiments~\cite{STAR:2014dcy,STAR:2018uho,Isshiki:2021bqh,CMS:2023jjt,ALICE:2011kmy,ALICE:2012aai,ALICE:2015hvw,ALICE:2015hav,ALICE:2019eol,ALICE:2018ysd,ALICE:2019gcn,ALICE:2019buq,ALICE:2020ibs,ALICE:2020mfd,ALICE:2022yyh}. In this technique, the relative momentum distribution of two hadrons in their center-of-mass frame is measured and compared with a reference distribution without interactions, allowing for the precise extraction of the correlation function that contains information for the hadron-hadron interactions~\cite{Lisa:2005dd,Zhang:2004db,Yu:2008vb,Xu:2024dnd,ALICE:2020wvi,ALICE:2020mkb,ALICE:2021njx}. The correlation functions can be also used to study the nature of exotic hadrons~\cite{Liu:2024uxn}.

The full potential of the femtoscopic correlation functions has not yet been fully exploited, as most studies are focused on the $s$-wave interaction~\cite{Liu:2025eqw,Molina:2023jov,Liu:2023wfo,Liu:2023uly,Albaladejo:2023wmv} and few works investigate the correlation function in higher partial waves using the Schrödinger equation~\cite{Murase:2024ssm,Murase:2025nlo,Sarti:2025sdo}. In fact, there are many baryon excited states that couple to their decay channels predominantly via higher partial waves~\cite{ParticleDataGroup:2024cfk}. The contributions from higher partial waves to correlation functions are often described using a simple Breit-Wigner parameterization~\cite{ALICE:2023wjz,Sarti:2023wlg,Kamiya:2019uiw}. However, this approach is inadequate for fully utilizing high-precision correlation function data to constrain the properties of some high-spin baryon states. Therefore, the construction of a universal theoretical framework that systematically incorporates contributions from arbitrary partial wave interactions is important and highly desirable.

In Ref.~\cite{Aceti:2012dd}, an analytical formalism for the partial-wave scattering matrix and wave functions describing hadron-pair interactions within a coupled-channel framework was presented using quantum mechanics. Based on the findings of Refs.~\cite{Aceti:2012dd,Gamermann:2009uq,Yamagata-Sekihara:2010kpd,Vidana:2023olz,Aceti:2014oma}, we systematically derive the analytic expression for the femtoscopic correlation function of hadron-hadron interaction through the Lippmann-Schwinger equation, which is applicable to all partial wave interactions. It is found that the obtained expression clearly reveals how different partial wave amplitudes contribute to the correlation function. As a first concrete application of this theoretical framework, we focus on the $d$-wave $K^- p$ interactions, in which the $\Lambda(1520)$ resonance with spin-parity $J^P =3/2^-$ is generated dynamically.

Indeed, the dynamically generated nature of $\Lambda(1520)$ resonance was investigated in Refs.~\cite{Aceti:2014wka,Roca:2006sz,Roca:2006pu,Sarkar:2005ap} within a chiral unitary approach involving the coupled channel interactions of $\pi \Sigma(1385)$ and $K \Xi(1530)$ in $s$-wave, and $\pi \Sigma$ and $\bar{K} N$ in $d$-wave, and it was found that the $\Lambda(1520)$ resonance is essentially dynamically generated from these meson-baryon channels. The work in Ref.~\cite{Encarnacion:2024jge} studied the femtoscopic correlation functions of strangeness $S=-1$ meson-baryon pairs, using unitarized $s$-wave amplitudes based on a next-to-leading-order chiral Lagrangian. It successfully reproduced most features of the $K^- p$ correlation functions data but failed to reproduce the bump structure around $p_{K^-} = 240$ MeV, which is attributed to the $\Lambda(1520)$ resonance decaying into $K^- p$ in $d$-wave. 

Following this line of research, we adopt the interaction potential described in Refs.~\cite{Aceti:2014wka,Roca:2006sz,Roca:2006pu,Sarkar:2005ap}, which naturally includes the mechanism that dynamically generates the $\Lambda(1520)$. Using this formalism, we successfully compute the impact of the $d$-wave interaction on the $K^- p$ correlation function. Furthermore, the properties of the $\Lambda(1520)$ are extracted by fitting the $K^- p$ correlation function and the $D_{03}$ scattering amplitudes of $\bar K N \to \bar K N$ and $\bar K N \to \pi\Sigma$ processes. We find that the obtained pole position of $\Lambda(1520)$ are consistent with the experimental results quoted in the Particle Data Group (PDG)~\cite{ParticleDataGroup:2024cfk}.

This article is organized as follows. In Sec.~\ref{sec:wave function}, the analytical expression of the wave function for arbitrary partial waves is introduced. Sec.~\ref{sec:coulomb} presents how the Coulomb interaction is incorporated. The formalism for the $d$-wave interaction in the $K^-p$ system is described in Sec.~\ref{sec:interaction formalism}, along with the approach for calculating the properties of a dynamically generated resonance. The analytical expression for the correlation function applicable to any partial waves is derived in Sec.~\ref{sec:correlation function}. The numerical results and discussions are presented in Sec.~\ref{sec:Results}. A brief summary is provided in Sec.~\ref{sec:Summary and Conclusions}.


\section{Formalism} \label{sec:Formalism}

\subsection{Wave function for general partial waves in the coupled channels} \label{sec:wave function}

Following the formalism in Refs.~\cite{Aceti:2012dd,Gamermann:2009uq,Yamagata-Sekihara:2010kpd,Vidana:2023olz,Aceti:2014oma}, the derivation of the wave function for arbitrary partial waves in coupled channels is briefly introduced. As in Ref.~\cite{Aceti:2014oma}, a separable interaction potential matrix $V$ is adopted. The elements of the potential matrix can be expressed as
\begin{equation}
    \begin{aligned}
        \bkthree{q}{V_{ij}}{p}=&(2l_{ij}+1)P_{l_{ij}}(\cos\theta_{\hat{q}\hat{p}})V_{ij}(\vec q,\vec p)\\
        &\cuti{i}{q}\cuti{j}{p},
        \label{equ:Vij_lij}
    \end{aligned}
\end{equation}
where $V_{ij}(\vec{q},\vec p)=v_{ij}\pli{ii}{q}\pli{jj}{p}$ and $l_{ij}$ represents the orbital angular momentum between channel $i$ and channel $j$. $\vec{q}$ and $q$ are the three momenta and its corresponding magnitude. $P_{l_{ij}}(\cos\theta_{\hat{q}\hat{p}})$ is the Legendre polynomial for the partial wave $l_{ij}$th order. $\Lambda_i$ is a cutoff parameter and $\cuti{i}{p}$ is a step function as the modulating factor. The specific form of the modulating factor, whether it is a step function or an exponential function, does not affect the derivation and form of the following formulas \cite{Gamermann:2009uq,Yamagata-Sekihara:2010kpd,Aceti:2012dd}. 

According to the Lippmann-Schwinger equation,
\begin{equation}
    \begin{aligned}
        T=V+VGT=V+VGV+...,
    \end{aligned}
\end{equation}
the elements of scattering amplitude matrix $T$ in momentum space can be calculated,
\begin{equation}
    \begin{aligned}
        \bkthree{q}{T_{ij}}{p}=&(2l_{ij}+1){P_l}_{ij}(\cos\theta_{\hat{q}\hat{p}})T_{ij}(\vec{q},\vec p)\\
        &\cuti{i}{q}\cuti{j}{p},
        \label{equ:T_single}
    \end{aligned}
\end{equation}
where $T_{ij}(\vec{q},\vec p) = t_{ij}\pli{ii}{q}\pli{jj}{p}$. The two-body scattering amplitude $t_{ij}$ can then be calculated algebraically,
\begin{equation}
    \begin{aligned}
        t=\frac{v}{1-v \tilde{G}},
        \label{equ:T_algebraical}
    \end{aligned}
\end{equation}
where $v$ is the two-body transition potential, and it does not contain explicitly the momenta $p^{l_{ij}}$, which is absorbed into the definition of the loop function $\tilde G$ for the general waves,
\begin{equation}
    \begin{aligned}
        \tilde{G}_{ii} (s) =\int^{\Lambda_i}_0 \rmd^3k~G_{ii}(k) k^{2l_{ii}},
        \label{equ:dotG}
    \end{aligned}
\end{equation}
with
\begin{equation}
    \begin{aligned}
        G_{ii}(k)=\frac{2 M_i}{(2\pi)^3}  \frac{(\omega_{M_i}+\omega_{m_i})}{2\omega_{M_i} \omega_{m_i}\left[s-(\omega_{M_i}+\omega_{m_i})^2+\rmi\epsilon\right]},
    \end{aligned}
\end{equation}
where $m_i$, $M_i$, $\omega_{M_i}=\sqrt{M_i^2+k^2}$, and $s$ are the meson mass, the baryon mass, the baryon energy, and the invariant mass squared of the $i$th meson-baryon pair, respectively.

The scattering wave function $\Psi$ in momentum space in the center-of-mass frame of the hadron-hadron pair can be obtained by the free relative wave function $\phi$, the loop function $G$, and the scattering amplitude $T$,
\begin{equation}
    \begin{aligned}
        \Psi=\phi+GT\phi,
    \end{aligned}
\end{equation}
where $\phi = \ket{\vec p}$ is the momentum eigenstate. To express the wave function in coordinate space, a complete set of momentum eigenstates is inserted, and it can be obtained by the Fourier transform, adopting the normalization conventions as done in Refs.~\cite{Aceti:2012dd,Vidana:2023olz},
\begin{eqnarray}
  &&      \int \rmd^3 p\equiv \ket{\vec p}\bra{\vec p},\quad \braket{\vec q|\vec p}=\delta^3(\vec p-\vec q), \\
  &&      \braket{\vec r|\vec p}=\frac{e^{\rmi \vec p \cdot \vec r}}{(2\pi)^{3/2}},\quad \tilde \Psi=(2\pi)^{3/2}\Psi. 
\end{eqnarray}

Therefore, the scattering wave function for the $i$th channel in coordinate space can be written as,
\begin{eqnarray}
\tilde\Psi_i(\vec r) &=& \tilde\phi_i(\vec r) + \nonumber \\
&& \sum_k \rmi^{l_{ik}}(2l_{ik}+1){P_l}_{ik}(\cos\theta_{\hat{p_k}\hat{r}})R_{ik}(\vec p_k),
        \label{equ:wavefunction_coordinate}
\end{eqnarray}
with the radial function
\begin{equation}
        R_{ik}(\vec p_k)=\int^{\Lambda_i}_0 \rmd^3q~j_{l_{ik}}(qr)G_{ii}(q)T_{ik}(\vec q,\vec{p}_k),
        \label{equ:radial-function}
\end{equation}
where $j_{l_{ik}}(qr)$ is the $l_{ik}$th order spherical Bessel function.

\subsection{Coulomb interaction} \label{sec:coulomb}

In the case where both hadrons are charged, the Coulomb interaction must be taken into account in addition to the strong interaction. In this paper, the approach of Refs.~\cite{Encarnacion:2024jge,Torres-Rincon:2023qll,Kamiya:2019uiw} is employed for the calculation of the Coulomb interaction. First, the Coulomb potential in coordinate space can be Fourier-transformed into momentum space,
\begin{eqnarray}
\!\!\!\!\!\!      V^{\mathrm{c}}(q) = - \int_0^{R_c}\rmd^3r~\mathrm{e}^{i\vec q \cdot  \vec r}~\frac{\alpha}{r} = - \frac{4\pi\alpha}{q^2}[1-\cos(qR_c)],
\end{eqnarray}
where $q=|\vec p-\vec p'|$ is the magnitude of the relative momentum, $\alpha=1/137$ denotes the fine structure constant. The cutoff parameter $R_c$ is introduced to render the Fourier transform numerically tractable. In this work, we take $R_c = 60$ fm, as used in Ref.~\cite{Encarnacion:2024jge}.

Subsequently, in order to obtain the wave function in Eq.~(\ref{equ:wavefunction_coordinate}) and the total scattering amplitude within a specific partial wave, a partial wave expansion of the Coulomb potential is required. The expansion is performed via
\begin{equation}
    \begin{aligned}
        V^{c}_l&=\frac{1}{2}\int^{1}_{-1}\rmd\cos\theta~V^c(q)P_l(\cos\theta)\\
        &=-\frac{1}{2}\int^{q_{\text{max}}}_{q_{\text{min}}}\frac{\rmd q}{p p'}~V^c(q)P_l(q),
    \end{aligned}
\end{equation}
where $q_{\text{min}}=|p-p'|$ and $q_{\text{max}}=p+p'$. Here we use the relation $q^2=p^2+p'^{2}-2pp'\cos\theta$ to transform the integral variable from $\cos\theta$ to $q$. Since this Coulomb potential is non-relativistic, a relativistic correction factor is introduced to match the relativistic strong interaction amplitude:
\begin{equation}
    \begin{aligned}
        T^{c}_l(p,p') &=\sqrt{\frac{2\omega_M(p)2\omega_m(p)\xi(p)}{2M}} V^{c}_l  \\ 
        & \times \sqrt{\frac{2\omega_M(p')2\omega_m(p')\xi(p')}{2M}},
    \end{aligned}
\end{equation}
where
\begin{equation}
    \begin{aligned}
        \xi(p)=\frac{\sqrt{s}-\omega_M(p)-\omega_m(p)}{p_{\text{on}}^2/2\mu-p^2/2\mu},
    \end{aligned}
\end{equation}
where $p_{\text{on}}=\lambda^{1/2}(s,M^2,m^2)/(2\sqrt{s})$ is the on-shell three-momentum, and $\lambda(a,b,c)=a^2+b^2+c^2-2ab-2ac-2bc$. 

Finally, the total scattering amplitude in Eq.~(\ref{equ:radial-function}), incorporating both strong and Coulomb interactions, is then obtained as
\begin{equation}
    \begin{aligned}
        T_{ik}(\vec q,\vec p_k)=t_{ik} q^{l_{ii}} p_k^{l_{kk}}+\delta_{ik} T_{l_{ik}}^c(q,p_k).
        \label{equ:T_tot}
    \end{aligned}
\end{equation}

\subsection{The coupled channels formalism for the $K^-p$ system and the properties of the $\Lambda(1520)$} \label{sec:interaction formalism}

In this section, the coupled channels formalism for the $K^-p$ $s$- and $d$-wave interactions is introduced and the properties of the $\Lambda(1520)$ are extracted from the interactions of the $\pi\Sigma(1385)$ and $K\Xi(1530)$ in $s$-wave and $\bar K N$ and $\pi \Sigma$ in $d$-wave. 

As done in Refs.~\cite{Aceti:2014wka,Roca:2006sz,Roca:2006pu,Sarkar:2005ap}, the $d$-wave interaction potential matrix is constructed, which includes the isospin $I=0$ channels, such as $\pi\Sigma(1385)$ and $K\Xi(1530)$ in $s$-wave and $\bar K N$ and $\pi \Sigma$ in $d$-wave. It is given by
\begin{equation}    
    \begin{aligned}
    v=
    \begin{pmatrix}
    C_{11}\left(k_1^0+k_1^0\right)&C_{12}\left(k_1^0+k_2^0\right)&\gamma_{13}&\gamma_{14}\\
    C_{12}\left(k_1^0+k_2^0\right)&C_{22}\left(k_2^0+k_2^0\right)&0&0\\
    \gamma_{13}&0&\gamma_{33}&\gamma_{34}\\
    \gamma_{14}&0&\gamma_{34}&\gamma_{44}
    \end{pmatrix},
    \label{equ:Vd}
\end{aligned}
\end{equation}
where the indices $1$ to $4$ are assigned to the channels $\pi\Sigma(1385)$, $K\Xi(1530)$, $\bar K N$, and $\pi \Sigma$, respectively. The coefficients of pseudoscalar mesons-baryon decuplet interaction are $C_{11}=-1/f^2$, $C_{12}=C_{21}=-\sqrt{6}/(4f^2)$, and $C_{22}=-3/(4f^2)$, where $f=1.15\times93\text{ MeV}$ is the average meson decay constant. The $k_i^0$ denotes the meson energy in the $i$th channel:
\begin{eqnarray}
k_i^0 &=&  \frac{s+m_i^2-M_i^2}{2\sqrt{s}},
\end{eqnarray}
where $\sqrt{s}$, $m_i$, and $M_i$ are the center-of-mass energy, the meson mass, and the baryon mass in the $i$th channel.

The $s$-wave interaction of $K^-p$ and the corresponding dynamical parameters in coupled channels are taken from Ref.~\cite{Oset:2001cn}. The $d$-wave dynamical parameters $\gamma_{ij}$ ($i=1$, $3$ and $j=3$, $4$) in Eq.~(\ref{equ:Vd}) and the cutoff parameters $\Lambda$ in Eq.~(\ref{equ:dotG}) are treated as model parameters, and their values are determined by fitting them to the experimental data on the $K^- p$ correlation functions and $d$-wave interaction amplitude, which will be discussed below. For partial waves other than the $s$- and $d$-waves, we neglect any dynamic interactions and treat them as background contributions.

The loop function for stable particles is defined as in Eq.~(\ref{equ:dotG}). However, since the $\Sigma(1385)$ is an unstable resonance and the mass threshold of $\pi\Sigma(1385)$ is close to the nominal mass of $\Lambda(1520)$ state, we modify the $\pi \Sigma(1385)$ loop function including the effect of the width of $\Sigma(1385)$ by the spectral function of the $\Sigma(1385)$. This treatment follows the methodology as done in Refs.~\cite{Lu:2020ste,Xie:2007qt}:
\begin{eqnarray}
\bar G_{11} (s) &=&  \frac{\int^{\tilde M_{\text{max}}^2}_{\tilde M_{\text{min}}^2} \rmd\tilde M^2~ \tilde G_{11}(s,\tilde M^2)~N(\tilde M^2)}{\int^{\tilde M_{\text{max}}^2}_{\tilde M_{\text{min}}^2} \rmd\tilde M^2~ N(\tilde M^2)},
\end{eqnarray}
with
\begin{eqnarray}    
    N(\tilde M^2)=-\frac{1}{\pi}\text{Im}\left(\frac{1}{\tilde M^2-M_{\Sigma(1385)}^2+iM_{\Sigma(1385) }\tilde\Gamma_{\Sigma(1385)}}\right ), \nonumber
\end{eqnarray}
where $\tilde M_{\text{min}}=M_{\Sigma(1385)} - 6 \Gamma_{\Sigma(1385)}$ and $\tilde M_{\text{max}}=M_{\Sigma(1385)} + 6 \Gamma_{\Sigma(1385)}$. Since the $\Sigma(1385)$ primarily decays into the $\pi\Lambda$ and $\pi\Sigma$ channels, the energy dependent width of $\Sigma(1385)$ is constructed to be dependent on these two decay channels. Therefore, the energy dependent width $\tilde\Gamma_{\Sigma(1385)}$ of $\Sigma(1385)$ is written as
\begin{eqnarray}    
    \tilde \Gamma_{\Sigma(1385)}(\tilde M^2) &=& \Gamma_{\Sigma(1385)}\frac{M_{\Sigma(1385)}}{\tilde M}\left[ 0.870\frac{p^3_{\pi\Lambda}(\tilde M^2)}{p^3_{\pi\Lambda}(M_{\Sigma(1385)}^2)}\right. \nonumber \\
    &&\left.\qquad+0.117\frac{p^3_{\pi\Sigma}(\tilde M^2)}{p^3_{\pi\Sigma}(M_{\Sigma(1385)}^2)}\right ],
\end{eqnarray}
where $p_{\pi\Lambda(\Sigma)}(\tilde M^2)=\lambda^{1/2}(\tilde M^2,M_{\pi}^2,M_{\Lambda(\Sigma)}^2)/(2\tilde M)$. Here we take $M_{\Sigma(1385)}=1384.58$ MeV, $\Gamma_{\Sigma(1385)}=37.20$ MeV, $M_{\pi}=138.04$ MeV, $M_{\Lambda}=1115.68$ MeV, and $M_{\Sigma}=1193.15$ MeV. The branching fractions of $\Sigma(1385)$ to $\pi\Lambda$ and $\pi\Sigma$ channels are $87\%$ and $11.7\%$, respectively.

Once the scattering amplitude was obtained, then one can also look for poles of the scattering amplitude on the complex plane of $\sqrt{s}$. The pole, $s_p$, on the second Riemann sheet could be associated with the $\Lambda(1520)$ resonance:
\begin{equation}
    \begin{aligned}
        \left.\text{det}\left(1-v \tilde {G}^{II}\right)\right|_{s=s_{p}}=0.
\end{aligned}
\end{equation}
The mass and the width of $\Lambda(1520)$ can be obtained with $\sqrt{s_{p}} = M_{\Lambda(1520)} - i \Gamma_{\Lambda(1520)}/2$.

In order to obtain the properties of $\Lambda(1520)$ in the coupled channels, the propagator must be extrapolated to the complex plane~\cite{Liu:2025eqw,Li:2024rqb,Khemchandani:2016ftn}. The analytical continuation of the loop function is given by
\begin{equation}
    \begin{aligned}
        \tilde G^{II}_{ii}(s) & = \left\{\begin{array}{ll}
        \tilde  G_{ii}^{(I)}(s),  & \text{Re}[\sqrt{s}]<\left(m_{i}+M_{i}\right) \\
        \tilde  G_{ii}^{(II)}(s), & \text{Re}[\sqrt{s}] \geq\left(m_{i}+M_{i}\right)
    \end{array},\right.
    \label{equ:G-complex-plane}
    \end{aligned}
\end{equation}
where $\tilde G_{ii}^{(I)}(s)$ is the loop function as in Eq.~\eqref{equ:dotG} and 
\begin{equation}
    \begin{aligned}
    \tilde G_{ii}^{(II)}(s) = \tilde G_{ii}^{(I)}(s)+\rmi\frac{2M_i}{4\pi\sqrt{s}}p_i^{2l_{ii}+1}.
    \label{equ:G-second-riemann}
    \end{aligned}
\end{equation}
The imaginary part of the three momenta $p_i$ is required to be positive to ensure a correct analytic continuation. 

Close to the pole $s_{p}$, the couplings of $\Lambda(1520)$ to different coupled channels can be determined as
\begin{equation}
    \begin{aligned}
        g_ig_j=\lim_{s\to s_{p}}(\sqrt{s}-\sqrt{s_{p}})T^{II}_{ij},
\end{aligned}
\end{equation}
where $g_{i}$ is the coupling constant of the $\Lambda(1520)$ resonance to the $i$-th channel. Thus, by determining the residues of the scattering amplitude $T$ at the pole, one can obtain the couplings of the $\Lambda(1520)$ resonance to different channels, which are complex in
general.

Within the determined coupling constant of $\Lambda(1520)$ to $\bar{K} N$ and $\pi \Sigma$, then we can calculate its $d$-wave partial decay widths as~\cite{Lu:2020ste}
\begin{equation}
    \begin{aligned}
        \Gamma_{\Lambda(1520)\to i}=\frac{|g_i|^2 M_i p_i}{2\pi M_{\Lambda(1520)}}.
\end{aligned}
\end{equation}

Furthermore, the compositeness $X$, which indicates the probability to find the hadronic molecular component in the resonance, can be evaluated via the loop function from Eq.~(\ref{equ:G-complex-plane}) \cite{Aceti:2012dd,Yamagata-Sekihara:2010kpd,Aceti:2014oma}, 
\begin{equation}
    \begin{aligned}
        X_i=-\left.\frac{{g_i}^2}{p_i^{2l_{ii}}}\frac{\rmd \tilde G^{II}_{ii}}{\rmd \sqrt{s}}\right|_{s=s_{p}}.
\end{aligned}
\end{equation}

Finally, in order to obtain the correlation function for the $K^-p$ pair in the charged basis, the $d$-wave scattering amplitude $t$ calculated in the isospin basis must be transformed into the charge basis:
\begin{equation}
    \begin{aligned}
        t^{I=0}_{i,K^-p}=\frac{1}{\sqrt{2}}t^{I=0}_{i,\bar K N}.
    \end{aligned}
\end{equation}

\subsection{Femtoscopic correlation functions for general partial wave in coupled channels} \label{sec:correlation function}

With the wave function in Eq.~(\ref{equ:wavefunction_coordinate}) and the total scattering amplitude in Eq.~(\ref{equ:T_tot}), the correlation function for a pair of hadrons in coupled channels is defined theoretically as
\begin{equation}
    C_i(\vec{p}_i) =\sum_{S,L,J}\omega_{(S,L,J)}  \sum_j  \int \rmd^3r~ \omega_j S_j(\vec r)  |\tilde\Psi_j(\vec r)|^2,
    \label{equ:cf_total}
\end{equation}
where $\omega_{(S,L,J)}$ is the weight factor, which depends on the individual spins $S_1$ and $S_2$, the total spin $S$, the orbital angular momentum $L$, and the total angular momentum $J$ of the pair. The specific expression of $\omega_{(S,L,J)}$ is given by~\cite{Mihaylov:2018rva,Ge:2025put},
\begin{equation}
     \omega_{(S,L,J)}=\frac{2S+1}{(2S_1+1)(2S_2+1)}\frac{2J+1}{(2L+1)(2S+1)}.
\end{equation}
For the $K^-p$ system in this work, the spins are $S_1=0$, $S_2=1/2$, and $S=1/2$. In the case of the $d$-wave interaction, the allowed total angular momenta are $J=3/2$ and $J=5/2$, corresponding to the weights $\omega_{(1/2,2,3/2)}=2/5$ and $\omega_{(1/2,2,5/2)}=3/5$, respectively. $\omega_j$ is the production weight of the $j$th coupled channel pair generated in collisions, and
\begin{equation}
        S_j\left(\vec{r}\right)=\frac{1}{\left(4\pi R_j^2\right)^{3/2}}~\rme^{-\frac{r^2}{4R_j^2}},
\end{equation}
is the emitting source function that represents the relative distance distribution when the $j$th coupled channel pair is emitted. Given the independence of the particle pair production mechanism from the interaction partial waves, the production factors $\omega_j$ and source functions $R_j$ for all these partial waves are assumed to be identical to those of the $\bar{K}N$ and $\pi\Sigma$ in $s$-wave coupled channel interactions as in Ref.~\cite{Encarnacion:2024jge}. In addition, in the absence of experimental constraints on the $\pi\Sigma(1385)$ and $K\Xi(1530)$ channels, their production factors and source functions are set to $1$ and $1$ fm, respectively.

When calculating the correlation function, the wave function in Eq.~(\ref{equ:wavefunction_coordinate}) should be an outgoing wave function. Therefore, the scattering wave function of $j$th channel in Eq.~(\ref{equ:cf_total}) is written as~\cite{Encarnacion:2024jge,Torres-Rincon:2023qll,Liu:2022nec,Albaladejo:2024lam,Ikeno:2023ojl,Molina:2023jov,Haidenbauer:2018jvl},
\begin{equation}
    \begin{aligned}
        \tilde\Psi_j(\vec r)&=\delta_{ij}\tilde\phi_j(\vec r)+\rmi^{l_{ij}}(2l_{ij}+1){P_l}_{ij}(\cos\theta_{\hat{p_i}\hat{r}})R_{ij}(\vec p_i)\\
        &=\delta_{ij}\left[\tilde\phi_j(\vec r)-\tilde\phi_{j,l_{ij}}(\vec r)+\tilde\phi_{j,l_{ij}}(\vec r)\right]\\
        &\quad+\rmi^{l_{ij}}(2l_{ij}+1){P_{l_{ij}}}(\cos\theta_{\hat{p_i}\hat{r}}) R_{ij}(\vec p_i)\\
        &=\delta_{ij}\left[\tilde\phi_j(\vec r)-\tilde\phi_{j,l_{ij}}(\vec r)\right]+\rmi^{l_{ij}}(2l_{ij}+1)\\
        &\quad {P_{l_{ij}}}(\cos\theta_{\hat{p_i}\hat{r}})\left[\delta_{ij}j_{l_{ij}}(p_j r)+R_{ij}(\vec p_i)\right].
        \label{equ:}
    \end{aligned}
\end{equation}
where the partial wave matrix $l_{ij}$ for $s$- and $d$- wave $K^-p$ interactions are defined as $l_{ij,s}=0$ and 
\begin{equation}
    \begin{aligned}
        l_{ij,d} & = \left\{\begin{array}{ll}
        0, & i\le 2~\text{and}~j\le 2 \\
        2, & \text { else }
    \end{array}.\right.
    \label{equ:lijd}
    \end{aligned}
\end{equation}
The second term of Eq.~(\ref{equ:}) implies that the total scattering amplitude for a specific partial wave influences only the wave function of the corresponding partial wave. In the case where only the Coulomb interaction is included, this term corresponds to the Coulomb wave function of the $l_{ij}$ partial wave \cite{Encarnacion:2024jge,Torres-Rincon:2023qll,Kamiya:2019uiw}. On the other hand, for the other partial waves where interactions are neglected, the wave functions retain the form of free relative wave functions. Therefore, $\tilde \phi_{j}(\vec r)=\rme^{\rmi \vec p_j \vec r}$ and $\tilde \phi_{j,l_{ij}}(\vec r)=\rmi^{l_{ij}}(2l_{ij}+1){P_{l_{ij}}}(\cos\theta_{\hat{p_j}\hat{r}})j_{l_{ij}}(p_j r)$.


Consequently, the femtoscopic correlation function for general partial waves can be obtained as
\begin{equation}
    \begin{aligned}
    &C_i(\vec{p}_i)=\sum_{S,L,J}\omega_{(S,L,J)} \sum_j  \int \rmd^3r~ \omega_j S_j(\vec r)  \\
    &\left[ \delta_{ij}\left|\tilde\phi_j(\vec r)-\tilde\phi_{j,l_{ij}}(\vec r)\right|^2 + (2l_{ij}+1) \right. \\
    & \left. \times \left|\delta_{ij}j_{l_{ij}}(p_j r)+R_{ij}(\vec p_i)\right|^2 \right].
\end{aligned}
\end{equation}
With the two-body scattering amplitudes $T_{ij}$ obtained within the chiral unitary approach, one can get the values of $R_{ij}$ for $(\pi \Sigma(1385), K\Xi(1530), \bar{K} N, \pi\Sigma)\to\bar{K} N$ interactions in $d$-wave, and $(\bar{K} N, \pi\Sigma, \pi\Lambda, \eta\Lambda, \eta\Sigma, K\Xi)\to\bar K N$ interactions in $s$-wave. While for other waves, we take $R_{ij} = 0$. Besides, it is worth noting that when $L=2$ and $J=5/2$, the scattering amplitude $D_{05}$ is known to be nearly zero around $1520$ MeV~\cite{Zhang:2013cua}. Therefore, the contribution from $d$-wave with total angular momentum $J=5/2$ is not considered here and the corresponding $R_{ij}$ is set to zero.

\section{Numerical results and discussions} \label{sec:Results}

To investigate the $d$-wave $K^-p$ interaction and the corresponding correlation functions, we analyze the $K^-p$ correlation functions for $p_{K^-}$ from the threshold to 400 MeV, and the $D_{03}$ scattering amplitudes of $\bar{K} N \to \bar{K}N$ and $\bar{K}N \to \pi \Sigma$ in the energy range of 1470 to 1650 MeV. When calculating the scattering amplitude $D_{03}$, the normalization convention is consistent with that in Ref.~\cite{Aceti:2014wka}. In this work, the dynamical parameters for the $d$-wave interaction are treated as free parameters, while the $s$-wave parameters are fixed to the values reported in Ref.~\cite{Oset:2001cn}. The fit is performed on a total of $75$ data points with $7$ free parameters. The resulting best-fit parameters are:
\begin{equation}
    \begin{aligned}
    \Lambda_s &\equiv \Lambda_{\pi\Sigma(1385)} = \Lambda_{K\Xi(1530)} = 2221 \pm 213~{\rm MeV}, \\
\Lambda_d &\equiv \Lambda_{\bar K N}=\Lambda_{\pi \Sigma} = 933 \pm 16~{\rm MeV}, \\
\gamma_{13} &= -(0.54 \pm 0.07) \times 10^{-7}~{\rm MeV}^{-3}, \\
\gamma_{14} &=  (0.70 \pm 0.08) \times 10^{-7}~{\rm MeV}^{-3}, \\
\gamma_{33} &=  -(2.01 \pm 0.13) \times 10^{-13}~{\rm MeV}^{-5}, \\
\gamma_{34} &= -(0.29 \pm 0.14) \times 10^{-13}~{\rm MeV}^{-5}, \\
\gamma_{44} &= -(2.90 \pm 0.29) \times 10^{-13}~{\rm MeV}^{-5}.
\end{aligned}
\end{equation}
The obtained $\chi^2/{\rm d.o.f}$ is 1.67, which is reasonably small.

\begin{figure}[htbp]
    \centering
    \includegraphics[scale=0.45]{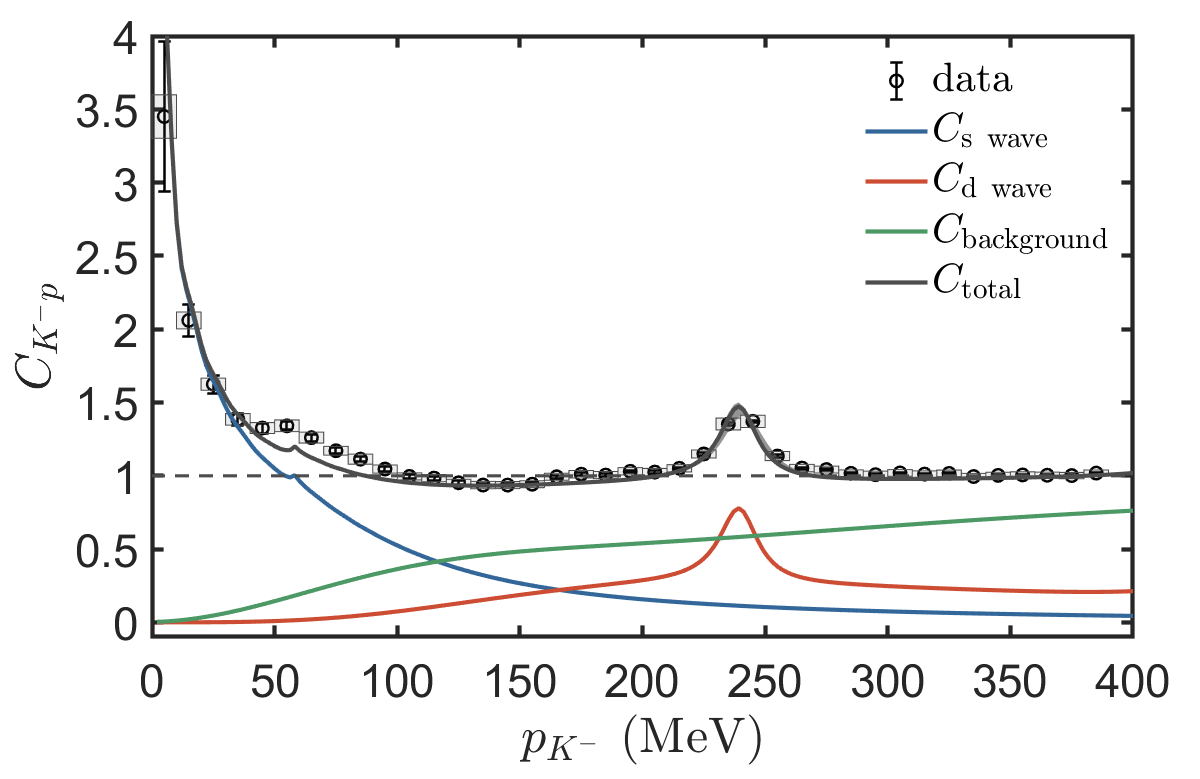}
    \caption{The fitted result of the $K^-p$ correlation functions. The experimental data are taken from Ref.~\cite{ALICE:2022yyh}. The blue and red curves represent the contributions from the $s$- and $d$-wave interactions, respectively. The green line denotes the background contribution from other partial waves. The black curve and gray shaded region indicate the total contribution from all partial waves and the corresponding $1\sigma$ uncertainty bands.}   
    \label{fig:cf_tot}
\end{figure}

First, we present the fitted results of the $K^- p$ correlation functions in Fig.~\ref{fig:cf_tot}, where the total contribution is depicted  by the dark grey curve. One can see that the experimental data are well reproduced. In Fig.~\ref{fig:cf_tot}, the contribution from $s$-wave $K^-p$ interaction is shown by the blue curve, while the $d$-wave contribution is shown by the red curve. The green line denotes the background contributions from the partial waves excluding the $s$- and $d$-waves, where the strong interactions are not considered. In the $K^-p$ correlation function, the result diverges when the relative momentum $p_{K^-}$ approaches zero, arising from the Coulomb interaction. A small structure is visible around $50~\text{MeV}$, which corresponds to the $\bar K^0 n$ mass threshold. This structure is identified as a manifestation of threshold effects between the $\bar K^0 n$ and $K^-p$ channels. It is worth noting that the theoretical curve lies slightly below the experimental data in this region. A more precise description of this specific region may require refitting the $s$-wave interaction parameters or including higher-order chiral Lagrangian to generate stronger attraction. However, this goes beyond the scope of the present work, which focuses on $d$-wave dynamics of $K^-p$ interaction, and does not affect our conclusions concerning the $\Lambda(1520)$ resonance. Furthermore, a pronounced peak is found near $240~\text{MeV}$, which is attributed to the $d$-wave $K^-p$ coupled channel interactions where the $\Lambda(1520)$ state is dynamically generated.

From Fig.~\ref{fig:cf_tot}, we observe that the $s$-wave contribution does not approach unity at larger momenta, unlike the results from other theoretical calculations~\cite{Liu:2024uxn,Shen:2025qpj} that consider only the $s$-wave contribution. The reason is that in those studies, contributions from other partial waves contained in the plane-wave are typically absorbed into the total correlation function. In our present framework, however, such contributions are explicitly separated and attributed to the 'background' term, which accounts for all other partial waves beyond the $s$-wave and also $d$-wave.

\begin{figure}[htbp]
    \centering
    \includegraphics[scale=0.45]{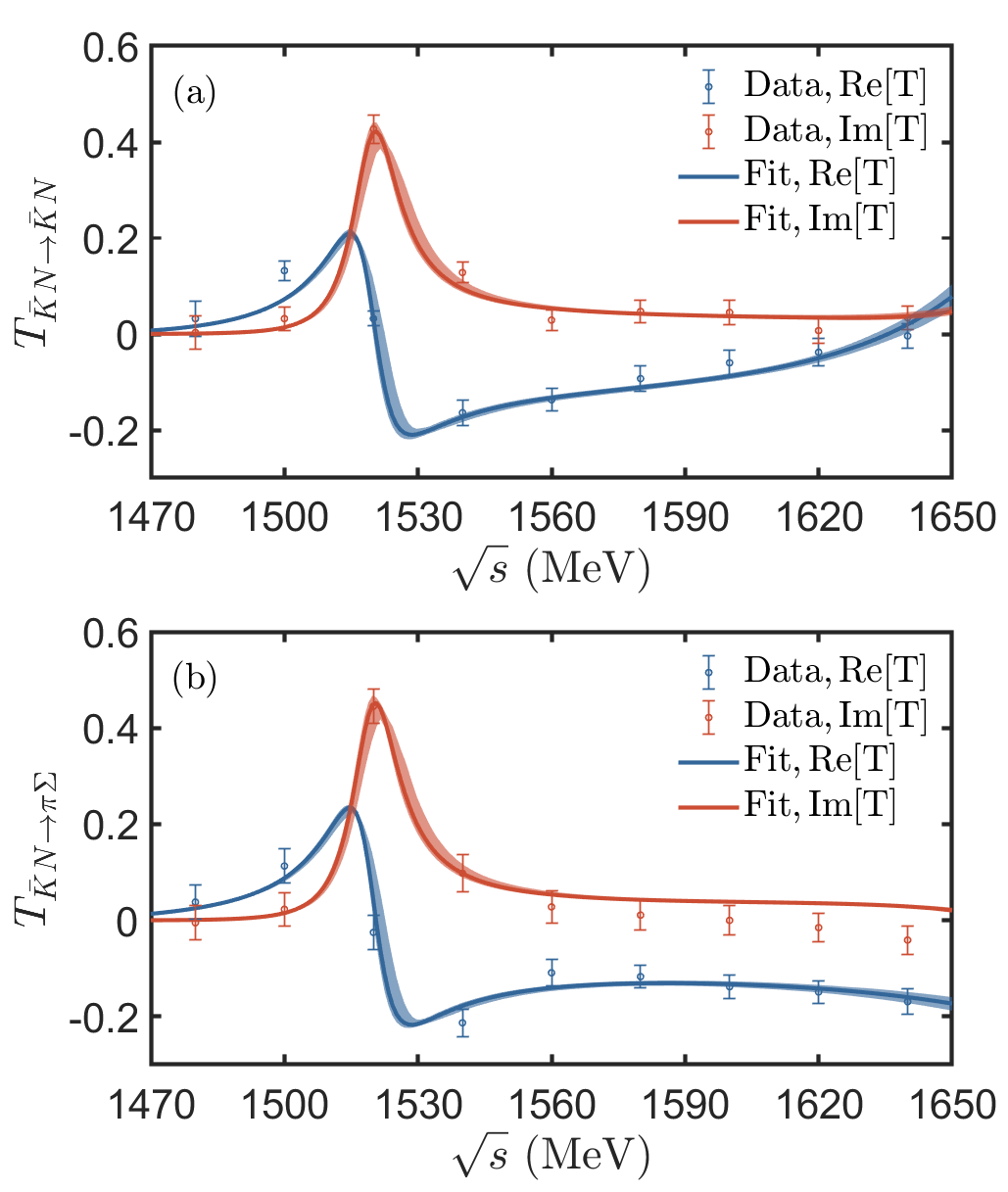}
    \caption{Fitted results of the $D_{03}$ scattering amplitudes $T_{\bar{K}N \to \bar{K}N}$ and $T_{\bar{K}N \to \pi\Sigma}$. The data are taken from Ref.~\cite{Zhang:2013cua}. The shaded regions represent the $1\sigma$ uncertainty bands of the  corresponding curves.}    
    \label{fig:D03}
\end{figure}

Second, the fitted results for the $D_{03}$ scattering amplitudes of $T_{\bar{K}N \to \bar{K}N}$ and $T_{\bar{K}N \to \pi\Sigma}$ are presented in Fig.~\ref{fig:D03} (a) and (b), respectively. It is found that we obtain a good description of the $D_{03}$ scattering amplitudes of $T_{\bar{K}N \to \bar{K}N}$ and $T_{\bar{K}N \to \pi\Sigma}$, especially for the shape of $\Lambda(1520)$ resonance.

Then, with these fitted model parameters, we can get the pole position of $\Lambda(1520)$, which is $\sqrt{s_{\rm pole}} = 1519.75 - i 6.57$ MeV. This result is consistent with the mass and width of $\Lambda(1520)$ quoted in the PDG~\cite{ParticleDataGroup:2024cfk}. Around the pole of $\Lambda(1520)$ resonance, its couplings and components to different channels and partial decay widths can be also obtained. These properties of $\Lambda(1520)$ are summarized in Table~\ref{tab:properties_Lambda(1520)}. From the obtained compositeness results, it can be concluded that the $\Lambda(1520)$ resonance has sizable $\bar K N$ and $\pi \Sigma$ molecular components, which is consistent with the conclusions of Ref.~\cite{Aceti:2014wka}. This indicates that the experimental data about the correlation functions can be used to study the exotic hadron states~\cite{Liu:2024uxn,Shen:2025qpj}.

\begin{table}[htbp]
\centering
\caption{The obtained properties for the $\Lambda(1520)$ resonance.}
\label{tab:properties_Lambda(1520)}
\begin{tblr}{
  width = \linewidth,
  colspec = {Q[180]Q[120]Q[210]},
  cells = {c},
  cell{1}{1} = {c=2}{0.308\linewidth},
  cell{2}{1} = {r=4}{},
  cell{6}{1} = {r=2}{},
  cell{8}{1} = {r=4}{},
  hline{1-2,6,8,12} = {-}{},
}
$\sqrt{s_{\rm pole}}$~ (MeV) &                  & $1519.75 - \text{i}~6.57$ \\
Coupling             & $g_{\pi \Sigma(1385)}$   & $0.69-\text{i}~0.05$      \\
                     & $g_{K\Xi(1520)}$   & $0.43-\text{i}~0.07$       \\
                     & $g_{\bar K N}$    & $0.47-\text{i}~0.07$     \\
                     & $g_{\pi\Sigma}$       & $0.43-\text{i}~0.06$       \\
{Partial Width\\(MeV)}     & $\Gamma_{\bar K N}$  & 5.24                                 \\
                     & $\Gamma_{\pi\Sigma}$  & 6.20                                 \\
{Compositeness} & $X_{\pi \Sigma(1385)}$   & $0.03$       \\
                     & $X_{K\Xi(1520)}$   & $0$        \\
                     & $X_{\bar K N}$    & $0.43$     \\
                     & $X_{\pi\Sigma}$        & $0.24$  
\end{tblr}
\end{table}

Next, we turn to the partial decay widths of $\Lambda(1520)$ to the $\bar{K}N$ and $\pi \Sigma$ channels in $d$-wave. Based on the extracted coupling constants $g_{\bar{K}N}$ and $g_{\pi\Sigma}$, branching ratios of $\Lambda(1520)$ to the $\bar{K}N$ and $\pi\Sigma$ channels can be calculated:

\begin{eqnarray}
Br[\Lambda(1520)\to\bar K N] &=& \frac{\Gamma_{\Lambda(1520)\to\bar K N}}{\Gamma_{\Lambda(1520)}} = 40\%, \\
Br[\Lambda(1520)\to \pi \Sigma] &=& \frac{\Gamma_{\Lambda(1520)\to \pi \Sigma}}{\Gamma_{\Lambda(1520)}} = 47\% ,
\end{eqnarray}
where we take $\Gamma_{\Lambda(1520)} = 13.14$ MeV. The theoretical results of the branching ratios are in agreement with the values listed in the PDG~\cite{ParticleDataGroup:2024cfk}.

\section{Summary and Conclusions} \label{sec:Summary and Conclusions}

In this work, we have established a generalized analytical framework for femtoscopic correlation functions incorporating arbitrary partial waves via the LS equation, which enables the precision studies of high-spin resonances in hadron-hadron interactions. By applying this framework to the $K^-p$ system, we conducted a combined analysis of $K^-p$ correlation function data and scattering amplitudes $D_{03}$ for $\bar{K}N \to \bar{K}N$ and $\bar{K}N \to \pi\Sigma$. In this analysis, we successfully reproduce the $K^-p$ correlation function data, specifically the peak structure around $240 \text{ MeV}$ that cannot be described in the $s$-wave $K^-p$ coupled channel interactions. This peak is associated with the $\Lambda(1520)$ which is dynamically generated in the $d$-wave $K^-p$ coupled channel interactions.

The combined analysis provides further constraints on the dynamical parameters of the $d$-wave $K^-p$ coupled channel interactions, from which the properties of $\Lambda(1520)$ are obtained. The extracted $\Lambda(1520)$ pole position is located at $1519.75 - i 6.57$ MeV and the derived branching ratios $Br[\Lambda(1520)\to\bar K N] = 40\%$ and $Br[\Lambda(1520)\to \pi \Sigma] = 47\%$ are in good agreement with their values quoted in the PDG~\cite{ParticleDataGroup:2024cfk}. Furthermore, the obtained compositeness of $\Lambda(1520)$ reveals large $\bar{K}N$ and $\pi\Sigma$ molecular components, consistent with the previous results of Ref.~\cite{Aceti:2014wka}.

\section*{Acknowledgments}

This work is partly supported by the National Key R\&D Program of China under Grant No. 2023YFA1606703, and by the National Natural Science Foundation of China under Grant Nos. 12575094, 12435007 and 12361141819.

\bibliography{reference}

\end{document}